# Combining EEG source connectivity and network similarity: Application to object categorization in the human brain

*Ahmad Mheich[1,2,3], Mahmoud Hassan[1,3], Olivier Dufor[4], Mohamad Khalil[2] and Fabrice Wendling[1,3]*

[1]INSERM, U1099, Rennes, F-35000, France
[2]AZM center-EDST, Lebanese University, Tripoli, Lebanon
[3]Université de Rennes 1, LTSI, F-35000, France
[4] Télécom Bretagne (Institut Mines-Télécom), UMR CNRS Lab-STICC, Brest, France

[1,3]firstname.name@univ-rennes1.fr
[2] firstname.name@ul.edu.lb
[4] firstname.name@telecom-bretagne.eu

**ABSTRACT**

A major challenge in cognitive neuroscience is to evaluate the ability of the human brain to categorize or group visual stimuli based on common features. This categorization process is very fast and occurs in few hundreds of millisecond time scale. However, an accurate tracking of the spatiotemporal dynamics of large-scale brain networks is still an unsolved issue. Here, we show the combination of recently developed method called 'dense-EEG source connectivity' to identify functional brain networks with excellent temporal and spatial resolutions and an algorithm, called SimNet, to compute brain networks similarity.
Two categories of visual stimuli were analysed in this study: immobile and mobile. Networks similarity was assessed within each category (intra-condition) and between categories (inter-condition). Results showed high similarity within each category and low similarity between the two categories. A significant difference between similarities computed in the intra and inter-conditions was observed at the period of 120-190ms supposed to be related to visual recognition and memory access. We speculate that these observations will be very helpful toward understanding the object categorization in the human brain from a network perspective.

*Index Terms*— Dense-EEG source connectivity, functional brain networks, object categorization, networks similarity.

## 1. INTRODUCTION

The brain is a complex network, composed of interconnected neuronal elements [1]. Many studies reported that the information processing in the human brain involves distant brain regions [2]. Considerable progress has been achieved in neuroimaging techniques that are able to record brain activity, such as Magneto/Electro encephalography (M/EEG) and functional magnetic resonance imaging (fMRI). The excellent temporal resolution of EEG (1ms) gives an advantage in comparison with fMRI. This feature is very crucial to track the brain activity in very short time periods as most cognitive tasks last less than a few hundred of milliseconds [3]. Recently we developed a new method called 'dense-EEG source connectivity' to identify functional brain networks with excellent temporal (1ms) and spatial resolutions [4, 5]. The method involves two main steps: i) the resolution of the EEG inverse problem to estimate the dynamics of the cortical sources and ii) the computation of functional connectivity to evaluate statistical relationships between the reconstructed sources.

Another challenge hold in the quantitative comparison of the networks related to various categories of stimuli [6]. In this context, the spatial location of the graph nodes (distributed over distinct cortical regions) is a key factor, often ignored by most of algorithms proposed previously. Here, we use a recently proposed algorithm called 'SimNet' for measuring similarity between graphs for which the coordinates of nodes are known [7]. The main feature of SimNet is to take into account the spatial location of nodes in order to find the similarity scores between the compared graphs.

In this paper we combine both 'dense-EEG source connectivity' method and 'SimNet' algorithm. Functional networks will be identified during visual recognition task including different semantic categories of images such as those analyzed in the presented paper: mobile and immobile. The similarity between networks within each category and between categories will be computed.

## 2. MATERIALS AND METHODS

### 2.1. Dense-EEG Data

Real data were collected from twenty healthy subjects with no neurological disease involved in a picture naming task. They were asked to name 120 pictures from different categories presented on a screen using E-Prime 2.0 software (Psychology Software Objects, Pittsburgh, PA). All pictures were shown as black drawings on a white background and selected from a database of 400 pictures standardized for French [8]. The cerebral activity was recorded using dense-EEG system (256 electrodes, EGI, Electrical Geodesic Inc.). The EEG signals were recorded with sampling frequency of 1 kHz and band-pass filtered between 30-45 Hz. Each trial was inspected visually and epochs contaminated by eye blinking or any other noise source were excluded from the analysis using the Brainstorm toolbox [9]. This study was approved by the National Ethics Committee for the Protection of Persons (CPP), (BrainGraph study, agreement number 2014-A01461- 46, promoter: Rennes University Hospital). We removed the electrodes located on the face as well as the electrodes showing high impedance. Overall, 180 (over 256) electrodes were retained as providing excellent quality signals over all subjects.

### 2.2. Dense-EEG source connectivity

The pipeline of the proposed method is illustrated in Figure 1 and includes different steps: i) solving the EEG inverse problem, ii) estimating the statistical dependencies (functional connectivity) between reconstructed sources iii) characterizing the identified networks (in the form of nodes connected by edges forming a graph) by graph theory based analysis and vi) segmenting, in time, the cognitive process as a sequence of functional connectivity states.

Recently, Hassan et al. [3] proved that the combination of the weighted Minimum Norm Estimate (wMNE) with the Phase Locking Value (PLV) using high resolution EEG as the connectivity method is the best combination among the tested combination to study the functional connectivity at source level in the context of picture naming task. This combination was used in the presented work. To track the dynamics of brain functional connectivity, we used an algorithm allowing the segmentation of cognitive task into functional connectivity states [4]. This algorithm is based on the K-means clustering of the connectivity networks. (See Hassan et al. [5] for detailed description of the method).

### 2.3. Network similarity algorithm

In order to compare the brain networks, we used a network similarity algorithm called 'SimNet' recently developed in our team [7]. SimNet is based on two main steps: i) *nodes distance*, in this part the algorithm includes three operations: insertion, deletion and substitution of nodes and ii) *edges distance*, it consists of calculating the sum of the weight difference between two edges of two compared graphs. The algorithm provides a normalized Similarity Index (SI): 0 for no similarity and 1 for two identical networks (same properties and topology). The algorithm was evaluated on simulated graphs and real applications and The algorithm is described in [7, 10].

### 2.4. Practical issues

Here we analyzed the networks identified at 120-190 ms of the cognitive process. This period results from an arbitrary choice guided by our previous results in a different dataset. It consists of the merging of two periods of time, from 120ms to 150 ms and from 151ms to 190 ms respectively. These windows correspond to visual perception and access to memory, as reported in [5] and in the general literature of object naming (see [2] for a review). The number of trials (presented visual stimuli) that generate a correct naming answer was 742 and 783 for mobile and immobile respectively. To analyze the similarity within each category: firstly, we select randomly two sets of 200 trials from the total number of trials. Secondly, the functional connectivity was computed using PLV method for each set. Finally, the similarity score between brain networks within each category is calculated between the selected sets (Figure 1). To analyze the inter-category similarity, we select randomly a group of 200 trials from each category. After computing the functional connectivity matrix for each group of trials, the similarity score is computed between brain networks for the two categories (Figure 1). These steps were repeated 50 times, in each iteration we select randomly the trials and then we calculate the similarity scores. Finally we get 50 similarity scores within each category and 50 inter-category similarity scores. Statistical test was performed using Wilcoxon test for $p<0.05$.

## 3. RESULTS

The results obtained from the application of SimNet algorithm on real brain networks are presented in Figure 2. Figure 2 -A- shows the boxplots of the similarity scores calculated during the period 120-190 ms which correspond to the visual perception and the access to memory process, as reported in [5]. The results show a high similarity values for mobile and immobile conditions and lower similarity values for immobile vs. mobile. The median values of similarity indexes were: 0.705, 0.72 and 0.66 for immobile, mobile, immobile vs. mobile, respectively. Interestingly, a statistical significant difference was detected between immobile and immobile vs mobile ($p<0.05$) and between mobile and immobile vs mobile ($p<0.05$) (Wilcoxon test, Figure 2 -A-). No significant difference was observed between the scores of each category.

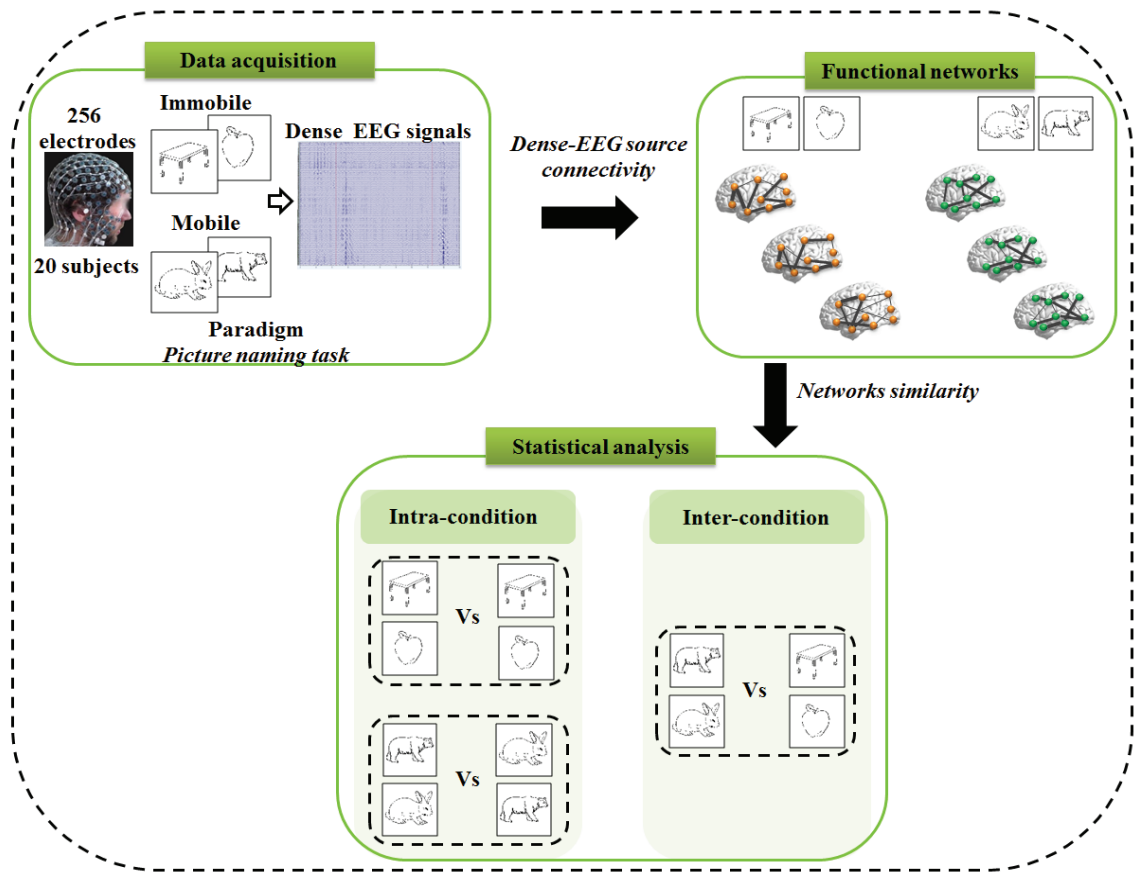

**Figure 1:** Structure of the investigation. Dense EEG (256 electrodes) were recording during picture naming task of different semantic categories including mobile and immobile, for 20 subjects. The functional connectivity was computed using 'Dense-EEG source connectivity' method. The similarity between the obtained networks was calculated using SimNet algorithm. This similarity was assessed within each category (intra-condition) and between categories (inter-condition).

Typical example of the brain networks for both categories of pictures (immobile and mobile) are illustrated in Figure 2-B- where the edges color represents their weights and the nodes size represents the strength value defined as the sum of weights of edges connected to this node.

## 4. DISCUSSIONS AND CONCLUSION

In this paper, we present a combination of 'dense EEG source connectivity method' and network similarity algorithm to investigate the object categorization in the human brain. The combination was applied to visual task consisting of naming pictures from different semantic categories. The results were very encouraging in the view of discriminating the networks with images from different semantic content. Results are briefly discussed hereafter.

First, the functional connectivity matrices using the phase locking value were computed at on all trials for all subjects, this could mask the inter-subject variability. This inter-subject variability was not explored here. However, it could be analyzed using the 'network-presence' index proposed in [5] when segmenting the visual task into functional connectivity states.

Second, in this paper, we focused on one time window related to the visual recognition and access to memory (120-190 ms). This window was chosen by merging second and third time windows originally identified by [5] on a different dataset. A natural perspective is the tracking of the similarity from the presentation of the stimulus to motor response. It is worth mentioning here that the similarity scores between brain networks were calculated without any thresholding on the connectivity matrices. However, a threshold was used only for visualization by taking the strongest 20% of edges.

Finally, to our knowledge this works has never been realized using dense EEG. The results are very encouraging and will help us to understand the object categorization in the human brain from a network-perspective.

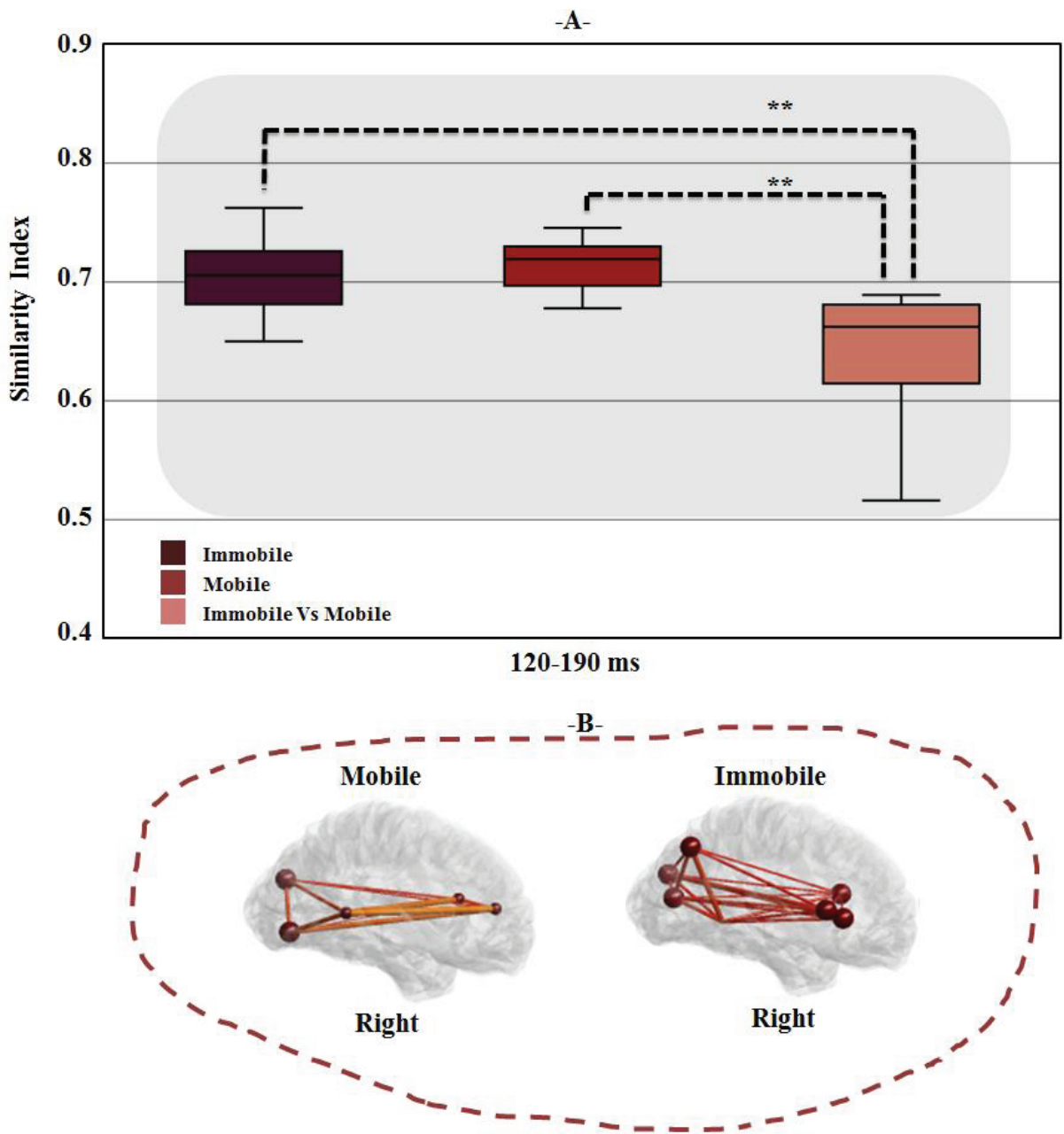

**Figure 2:** -A- Boxplots show significant difference of similarity values between intra/inter conditions during the period '120-190 ms of cognitive process using SimNet.-B- 3D representation of a typical example of networks for mobile and immobile conditions from the right view [11].

### ACKNOWLEDGEMENTS

This work was supported by AZM and SAADE Association (Tripoli, Lebanon) and the Rennes University Hospital (COREC Project named BrainGraph, 2015-17). The work has also received a French government support granted to the CominLabs excellence laboratory and managed by the National Research Agency in the "Investing for the Future" program under reference ANR-10-LABX-07-01. This work was also supported by the European Research Council under the European Union's Seventh Framework Programme (FP7/2007-2013) / ERC grant agreement n° 290901.